**A Kagome-Derived Mosaic Lattice Family $A_3V_9Te_{13}$ ($A$ = Cs, Rb) with Tunable Strong Electronic Correlations**

Yusen Xiao[1⊥], Zhibin Qiu[3,4⊥], Qingchen Duan[1], Zhaoyi Li[3,4], Xiaotong Xu[1], Hengxin Tan[2], Shu Guo[3,4*], and Ruidan Zhong[1,2*]

[1] Tsung-Dao Lee Institute, Shanghai Jiao Tong University, Shanghai 201210, People's Republic of China.

[2] School of Physics and Astronomy, Shanghai Jiao Tong University, Shanghai 200240, People's Republic of China.

[3] Shenzhen Institute for Quantum Science and Engineering, Southern University of Science and Technology, Shenzhen 518055, China.

[4] International Quantum Academy, Shenzhen 518048, China.

[*]Correspondence should be addressed to Shu Guo (guos@sustech.edu.cn) and Ruidan Zhong (rzhong@sjtu.edu.cn)

**Abstract**

The pursuit of geometrically frustrated lattices beyond conventional paradigms remains a central challenge in the design of quantum materials. Herein, we report the discovery of the $A_3V_9Te_{13}$ ($A$ = Cs, Rb) family of vanadium-based intermetallic compounds, which host a unique two-dimensional Mosaic lattice derived from the Kagome network, composed of an ordered tessellation of triangles, squares, and pentagons. The Cs compound (CVT) exhibits strong electronic correlations, characterized by non-Fermi liquid behavior at low temperatures, an exceptionally large Sommerfeld coefficient, and a bulk phase transition at $T^* \approx 47$ K with possible charge- or spin-related origin. Inspired by pressure-tuning in related Kagome systems, we demonstrate that the electronic ground state of this lattice is exquisitely tunable via chemical pressure. Systematic substitution of Cs with smaller Rb ions suppresses the $T^*$phase transition and the correlated electronic response, ultimately driving the system into a highly frustrated semiconducting ground state without long-range magnetic order down to 60 mK. This work unveils a new structural platform for exploring the interplay between geometric frustration and strong electron correlations, providing a chemically controllable platform for exploring the phase space between distinct correlated electronic states.

## 1. Introduction

The interplay between lattice geometry and electronic correlations is a cornerstone of modern condensed matter physics. Archimedean lattices[1,2], such as triangular, square, and honeycomb networks, have long served as fundamental playgrounds for emerging phenomena ranging from high-$T_c$ superconductivity[3,4] to Dirac fermions[5,6]. Among these, the Kagome lattice has recently ascended to prominence. Its geometry of corner-sharing triangles naturally combines geometric frustration[7], non-trivial band topology[8,9], and flat electronic bands[10], providing fertile ground for exotic quantum states. Recent discoveries of $AV_3Sb_5$ ($A$ = K, Rb, Cs) family have substantiated this potential, revealing unconventional charge density wave (CDW), potential time-reversal symmetry breaking (TRSB), pair density wave (PDW), and superconductivity[8,11–14]. In the related magnetic compound $CsCr_3Sb_5$, pressure suppresses density-wave order and induces superconductivity, accompanied by quantum critical behavior and enhanced quasiparticle mass renormalization[15], underscoring the intimate coupling between lattice geometry and electronic correlations. These advances inspire us to design novel Kagome-derived lattices via anion valence engineering, to explore geometric structures and correlated physics beyond conventional Archimedean tiling.

Despite these advances, the structural landscape of two-dimensional periodic lattices extends far beyond conventional Archimedean tiling. Regular pentagons cannot tile the plane without distortion[16,17], imposing fundamental geometric constraints on periodic crystalline frameworks. Ordered two-dimensional periodic lattices incorporating pentagonal motifs therefore remain exceptionally rare, and their influence on correlated electronic states is largely unexplored.

In this work, we report the discovery of the $A_3V_9Te_{13}$ ($A$ = Cs, Rb) family of vanadium-based intermetallic compounds, in which the two-dimensional vanadium layer forms an ordered tessellation of triangles, squares, and pentagons. We define this unique geometry as a "Kagome-derived Mosaic lattice". This architecture introduces a pentagon-integrated periodic framework that is structurally related to, yet distinct from, the Kagome network, thereby expanding the geometric design space of frustrated lattices. $Cs_3V_9Te_{13}$ (CVT) exhibits a phase transition of possible charge or spin origin at $T^* \approx 47$ K, non-Fermi liquid resistivity behavior, and a large Sommerfeld coefficient, placing it among strongly correlated electron systems. The correlated ground state is exquisitely tunable via chemical pressure: systematic substitution of Cs with smaller Rb ions completely suppresses the electronic ordering transition and quenches the heavy quasiparticle response. Upon full suppression of the ordered state, the system evolves into a non-magnetic, semiconducting state with extreme magnetic frustration and no long-range magnetic order down to 60 mK. These results position the $A_3V_9Te_{13}$ Mosaic lattice as a versatile structural platform for realizing and tuning strong electronic correlations and quantum disorder in a purely $d$-electron environment.

## 2. Results and Discussion

### 2.1 Crystal structure of the Mosaic lattice $Cs_3V_9Te_{13}$

A millimeter-sized, air-stable, high-quality single crystal of CVT was obtained via a self-flux method (**Figure 1e).** The crystal structure, determined by single-crystal X-ray diffraction (SC-XRD), crystallizes in the hexagonal crystal system with the centrosymmetric space group of $P$-62$m$ (No.189). Detailed crystallographic details are provided in **Table S1-2** and **Figure S1**. Powder X-ray diffraction (PXRD) measurements reveal (00L) diffraction peaks, confirming the single-phase nature of the sample with the $ab$-plane as the natural growth plane. Furthermore, Rietveld refinement of the polycrystalline sample synthesized via solid-state reaction further validates the structural reliability (**Figure S2**). Notably, upon cooling from 300 to 30 K, only a slight lattice variation was observed, with no indication of a change in crystal symmetry, providing compelling evidence for the absence of a crystal structural transition across the entire temperature range (**Figure S3, Table S1**). As a representative, the crystal structure refined at 300 K from SC-XRD was used for the following discussion. From a structural perspective, CVT adopts a V-based 2D layered architecture, with nonmagnetic Cs ions located between the layers and serving as spacers (**Figure 1a**). In the asymmetric unit of CVT, there are three unique Te sites (3*g*, 4*h*, 6*i*), one unique Cs site (3*f*), and two unique V sites (3*g*, 6*k*). Each Cs atom is coordinated by ten Te atoms, forming a $CsTe_{10}$ polyhedron with Cs–Te bond distances ranging from 3.83 to 4.13 Å (**Figure 1d**). The 2D Mosaic layer in CVT can be described as a network of two types of triangles, one type of square, and one type of pentagon (**Figures 1b and c**). Each V-based triangle connects to three squares and three pentagons via either edge-sharing or corner-sharing. Five-fold rotational symmetry is crystallographically forbidden. To accommodate such a unique geometry within a 2D plane, the lattice adopts a mixed tiling of triangles, squares, and pentagons. The V-V bond distances across the five edges of the pentagon vary from 3.02 to 3.19 Å, which distorts the pentagon away from a regular structure (**Figure S1**). The interatomic distances of V1-V1 and V2-V2 are 2.718 Å and 3.185 Å, respectively. The former is close to the short V-V bond in $CsV_3Sb_5$, which promotes strong 3*d* orbital overlap and large bandwidths, thus favoring an itinerant non-magnetic metallic state[14,18,19]. In contrast, the significantly elongated V2-V2 distance suppresses orbital overlap, favoring localized magnetic moments and enhanced electronic correlations. X-ray photoelectron spectroscopy (XPS) suggests a potential mixed-valence state of vanadium, with the V $2p_{3/2}$ peak corresponding to $V^{3+}$ and $V^{4+}$ (**Figure S4**). Given the itinerant nature of V1, these mixed valence features are likely associated predominantly with the V2 sublattice, suggesting site-internal charge disproportionation.

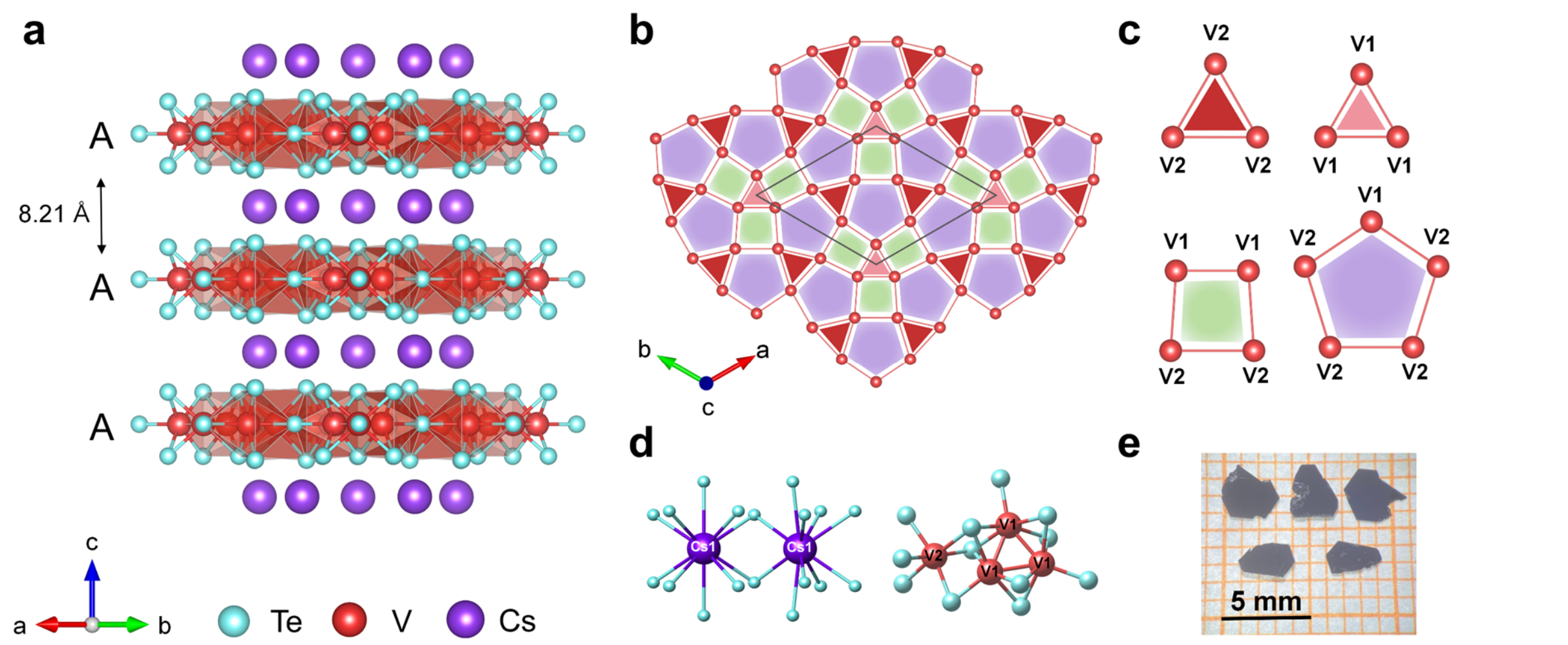

**Figure 1 Crystal structure. a**, Crystal structure of the Mosaic lattice compound CVT. **b**, 2D V-based Mosaic lattice, a tessellation of triangles, squares, and pentagons. The black rhombus outlines one unit cell. **c**, Basic structural units of the Mosaic lattice. **d**, Coordination environments for Cs and V atoms, respectively. **e**, Optical image of CVT single crystals.

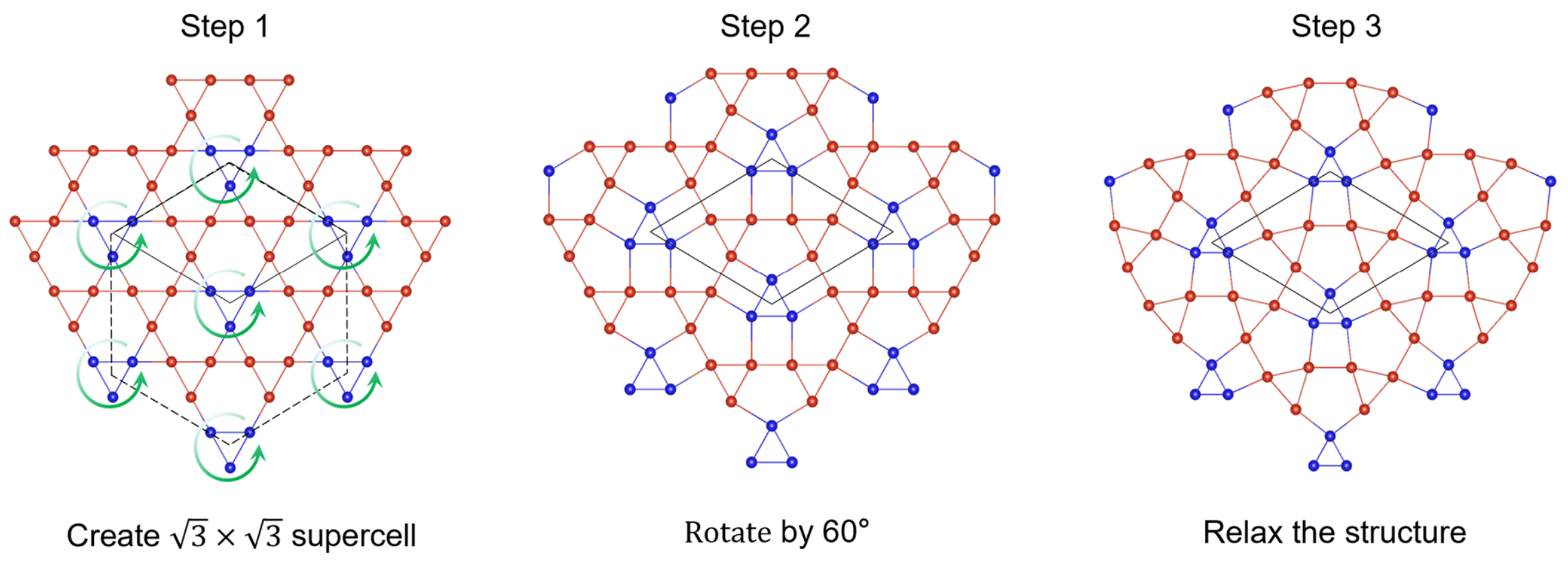

**Figure 2 Structural transformation from 2D Kagome to Mosaic lattice.** Left: The parent Kagome lattice, with the $\sqrt{3} \times \sqrt{3}$ supercell construction highlighted. Center: An idealized Mosaic lattice generated by 60° rotational reconstruction of the supercell. Right: The fully relaxed atomic configuration of the Mosaic lattice, which preserves the mixed triangle-square-pentagon tiling motif distinct from conventional Kagome geometry.

The geometry of a Mosaic lattice is fundamentally entwined with its electronic ground states. To elucidate the origin of this complex tiling, it is instructive to compare it with the prototypical Kagome lattice in $AV_3Sb_5$. Our Mosaic lattice could be described from a transformation product of an ideal Kagome lattice (**Figure 2**): first, atoms in the ideal Kagome lattice can be divided into two groups (blue triangular and the remaining red vanadium atoms) to create a $\sqrt{3} \times \sqrt{3}$ supercell (Step 1), and then all blue triangles are rotated by 60° (clockwise or counterclockwise) to generate a highly symmetric Mosaic lattice (Step 2). After full structural relaxation, the resulting geometry reproduces the Mosaic lattice observed in CVT (Step 3). Notably, this resulting Mosaic lattice retains the intrinsic geometric frustration of the parent Kagome network while introducing a higher degree of topological complexity. This structural metamorphosis not only expands the crystal unit cell, but also introduces two distinct V sites with drastically different V-V bond lengths, which goes beyond a simple renormalization of hopping parameters and significantly modify the electronic structure, providing a fertile platform for the strong electronic correlations and tunable quantum phases discussed hereafter.

### 2.2 Physical Properties of $Cs_3V_9Te_{13}$

The physical properties of the Kagome-derived Mosaic vanadium network were systematically investigated through comprehensive measurements (**Figure 3**). The temperature-dependent magnetic susceptibility of CVT single crystals, measured under an applied field of 1 T for both the *ab*-plane and the *c*-axis directions, exhibits a distinct kink at $T^* \approx 47$ K, signaling a phase transition (**Figure 3a**). The high-temperature data (50-150 K) follow the Curie-Weiss behavior (**Figure S5**). Fitting yields $\theta_{ab}$ = -62 K and $\theta_c$ = -37 K, indicating dominant antiferromagnetic interactions, with effective magnetic moments of $\mu_{\text{eff}} = 0.59\ \mu_B$/V and 0.56 $\mu_B$/V for $H$ // $ab$ and $H$ // $c$, respectively. Furthermore, the magnetization varies linearly with the magnetic field, without any sign of field-driven magnetic evolution (**Figure S6**). The experimentally determined effective moment is smaller than the spin-only values expected for V ions (e.g., 2.83 $\mu_B$ for $S$ = 1, 1.73$\mu_B$ for $S$ = 1/2), suggesting partial itinerancy of the V 3$d$ electrons. The relatively enhanced $\mu_{\text{eff}}$ compared with the Kagome metal $AV_3Sb_5$ (~0.2 $\mu_B$/V)[11] highlights the significant role of site-selective localization and electronic correlations within this Mosaic lattice.

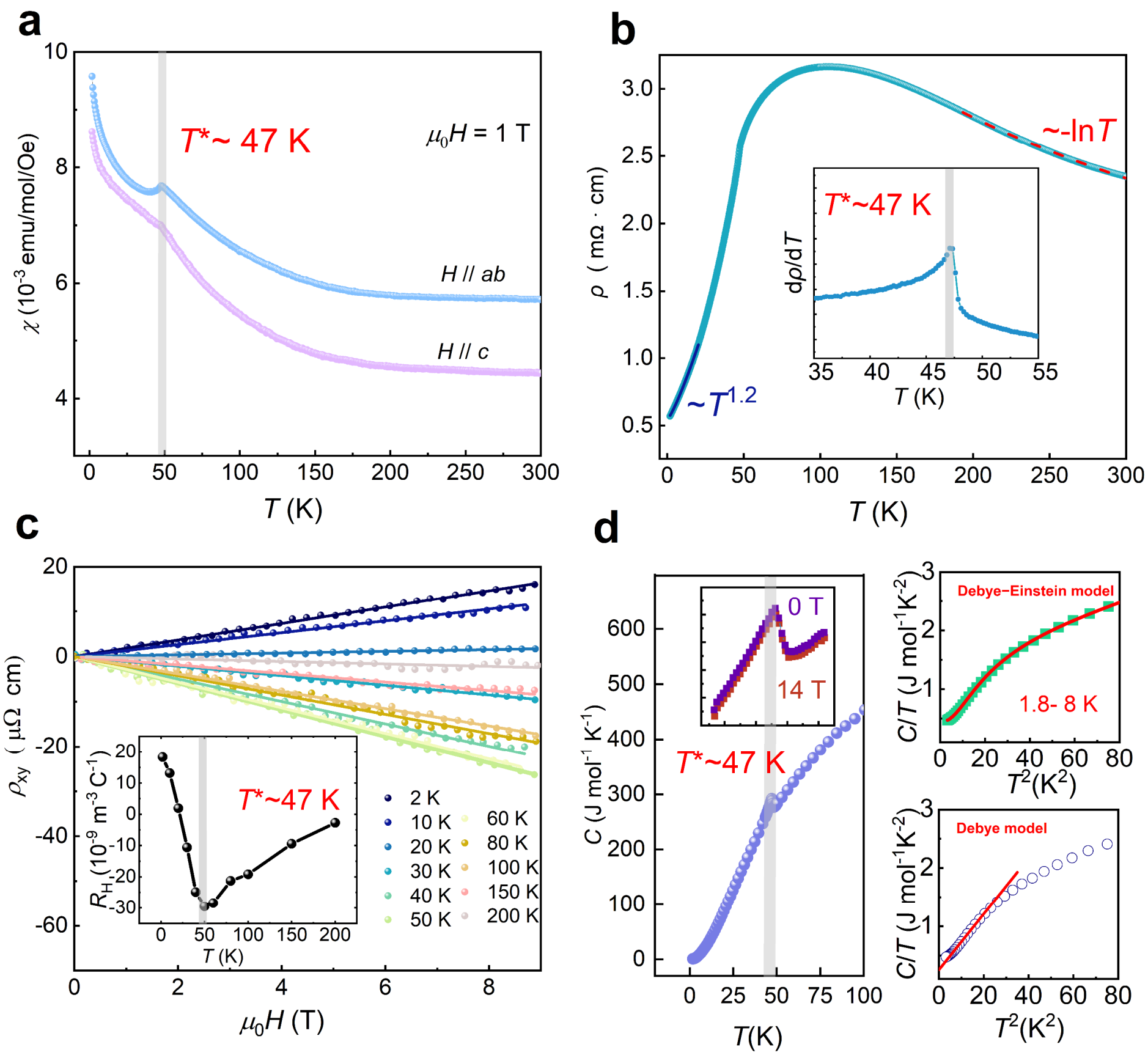

**Figure 3 Physical Properties of CVT. a**, Temperature-dependent magnetic susceptibility measured in an applied field of $\mu_0H$ = 1 T under zero-field-cooled (ZFC) conditions. **b**, Temperature-dependent resistivity $\rho(T)$, with the red dashed line representing the -ln$T$ trend and the blue solid line showing the fit using $\rho = \rho_0 + AT^n$. Inset of corresponding d$\rho$/d$T$ versus $T$ plot. **c**, Hall resistivity $\rho_{xy}$ at selected temperatures, determined via $\rho_{xy} = [\rho_{xy}(+H) - \rho_{xy}(-H)]/2$ to eliminate the longitudinal component. Inset of Temperature dependence of the Hall coefficient $R_H$. **d**, (Left) Temperature-dependent specific heat, with inset of measurements in the vicinity of $T^*$ under zero field and an applied field of 14 T, respectively, and (upper right) numerical fit to the data in the temperature range of 1.8-8 K using the Debye-Einstein model, (lower right) numerical fit to the data in the temperature range of 2.8-4.3 K using the Debye model.

The temperature-dependent electrical resistivity $\rho(T)$ of CVT shows characteristic behavior of a strongly correlated $3d$-electron system (**Figure 3b**). Above 150 K, $\rho(T)$ follows a logarithmic dependence $\rho \sim -\ln T$, consistent with incoherent scattering. Upon cooling, $\rho(T)$ reaches a broad maximum of ~3.1 mΩ·cm near 100 K, marking the onset of electronic coherence. A kink at $T^* \approx 47$ K is observed in the resistivity, coinciding with the anomaly identified in the magnetic susceptibility measurements. The transition exhibits negligible thermal hysteresis and remains insensitive to magnetic fields up to 5 T (**Figure S7**). These transport signatures are reminiscent of the electronically driven ordering transitions observed in kagome metals $AV_3Sb_5$[11,18,20]. At low temperatures, the resistivity follows a power-law form, $\rho = \rho_0 + AT^n$, with the exponent $n \approx 1.2$ in the range of 2– 20 K. This value deviates substantially from the Fermi-liquid expectation of $n = 2$, suggesting the presence of strong electronic correlations. The Hall resistivity $\rho_{xy}$ remains linear with magnetic field over the entire measured temperature range (**Figure 3c**). The extracted Hall coefficient $R_H$ displays a sharp drop at $T^* \approx$ 47 K, followed by a sign reversal below ~20 K, indicating hole-dominated conduction in the low-temperature regime. These anomalies suggest significant reconstruction of the electronic structure across $T^*$, likely involving changes in carrier density and/or mobility. Together with the resistivity anomaly, these results point to an electronically driven instability with possible charge- or spin-related origin.

Specific heat measurements provide direct thermodynamic insight into the nature of the phase transition (**Figure 3d**). A pronounced $\lambda$-type anomaly is observed at $T^* \approx 47$ K, without detectable thermal hysteresis (**Figure S8a**), indicating a second-order or weakly first-order phase transition. This is consistent with the absence of any detectable structural phase transition in SC-XRD measurements down to 30 K. The transition temperature remains essentially unchanged under magnetic fields up to 14 T, indicating that the ordered state is robust against magnetic fields. After subtracting the phonon background, the entropy change associated with the transition, obtained from $\Delta S = \int \Delta C/T dT$, reaches $\Delta S \approx 1.0$ J mol$^{-1}$ K$^{-1}$ (**Figure S8b**). The small entropy change is insufficient to be attributed to the full ordering of localized magnetic moments, suggesting instead an itinerant electronic origin with possible charge- or spin-related character. To extract the electronic specific-heat coefficient, we plot $C/T$ versus $T^2$ at low temperatures. The data deviate markedly from linearity, and we therefore employ a combined Debye-Einstein model[21] to describe the specific heat, fitting the data to

$$\frac{C_\mathrm{p}(T)}{T} = \gamma + \beta T^2 + 3R\delta \frac{(\Theta_\mathrm{E}/T)^2 e^{\Theta_\mathrm{E}/T}}{T[e^{\Theta_\mathrm{E}/T}-1]^2} ,$$

in which the first two terms originate from the standard Debye model, while the third term represents the contribution of an Einstein mode. The fit yields parameters $\gamma \approx$ 425 mJ mol$^{-1}$K$^{-2}$, $\beta$=14.1 mJ mol$^{-1}$ K$^{-4}$, and $\Theta_\mathrm{E}$ = 22.8 K. Notably, even though the pure Debye model (excluding the Einstein component) is applicable only over a limited temperature range , we still obtain a large Sommerfeld coefficient of ~256 mJ mol$^{-1}$ K$^{-2}$, confirming the presence of strong electronic correlations. The Sommerfeld coefficient $\gamma$ substantially exceeds values for the prototypical Kagome metal $CsV_3Sb_5$ ($\gamma$ ~ 20 mJ mol$^{-1}$ K$^{-2}$)[22] and $CsCr_3Sb_5$ ($\gamma$ ~ 105 mJ mol$^{-1}$ K$^{-2}$)[15], and is comparable in magnitude to that reported for $LiV_2O_4$ ($\gamma$ ~ 420 mJ mol$^{-1}$ K$^{-2}$), the rare *d*-electron heavy Fermion system driven by correlation-induced flat bands arising from intra-atomic Hund's coupling[23,24]. Such a large Sommerfeld coefficient indicates a pronounced enhancement of the electronic effective mass and strong electronic correlations within the Mosaic lattice. Collectively, the magnetic susceptibility, electrical transport, and specific heat measurements reveal a strongly correlated electronic state in CVT, while the transition at $T^*$~ 47 K appears to arise from an electronically driven instability with possible charge- or spin-related character.

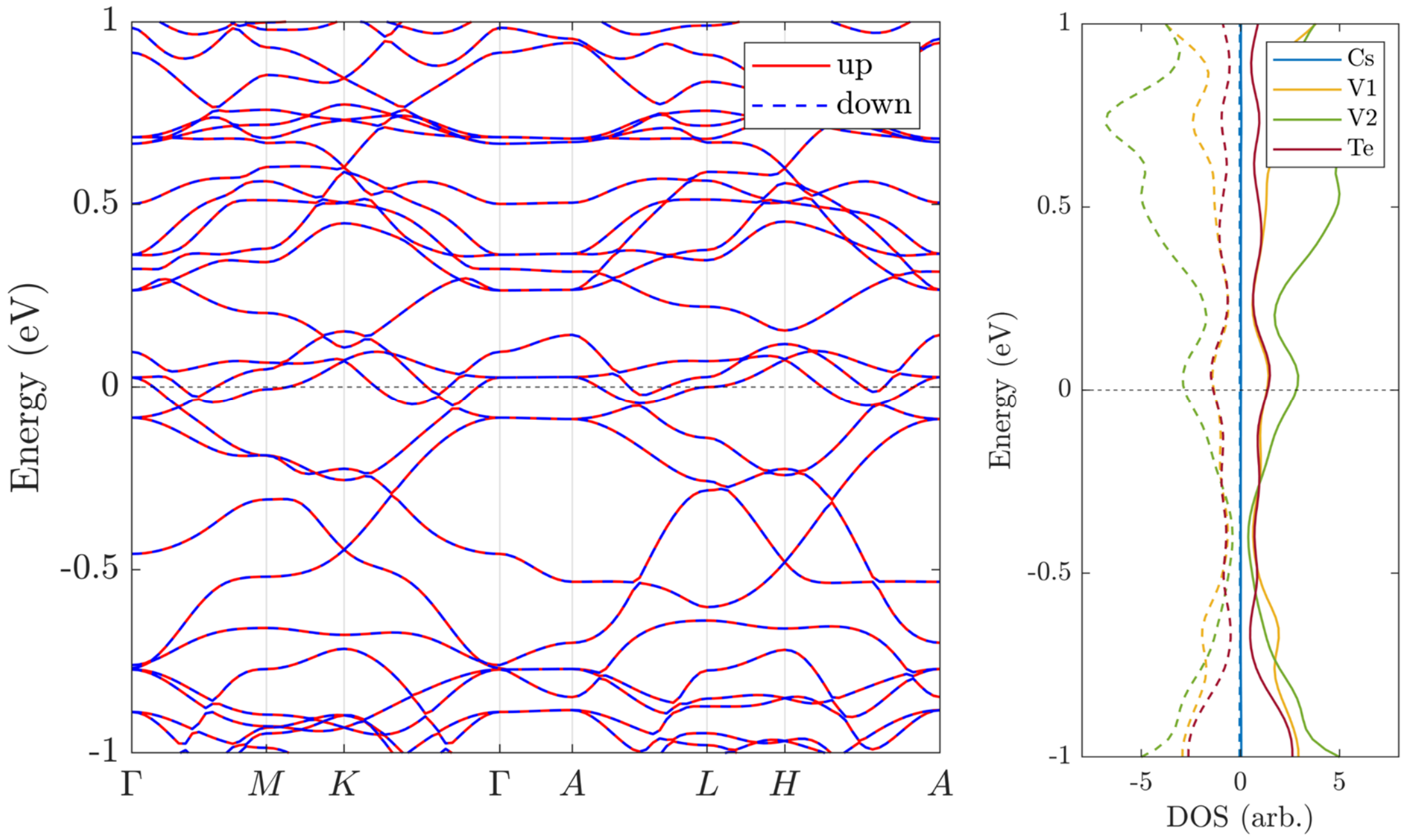

**Figure 4 First-principles electronic structure calculations of CVT.** Left: band structure where up and down represent the two spin channels. Right: density of states (DOS), position values for the spin up channel and negative ones for the spin down channel. Spin-orbit coupling is not included.

Electronic structure calculations provide crucial insight into the properties of CVT (**Figure 4**). Band structure and density of states (DOS) calculations were performed within an antiferromagnetic framework, with ferromagnetic alignment in the *ab* plane and antiferromagnetic coupling along the out-of-plane direction. The calculations yield local magnetic moments of ~0.1$\mu_B$ for V1 and ~1.57 $\mu_B$ for V2. The negligible local moment on V1 and sizable moment on V2 are consistent with their distinct local structure and electronic environments, and further support the mixed-valence character of the V2 sublattice, as implied by XPS analysis. Meanwhile, the electronic structure exhibits an overall metallic character, with multiple bands crossing the Fermi level. States near the Fermi level are predominantly derived from V 3*d* orbitals. The band structure hosts Dirac points near the K points of the Brillouin zone, van Hove singularities (vHSs) at the M points, and flat-band features within the Γ-M-K plane. These topological features are analogous to those recently reported in conventional Kagome superconductors $AV_3Sb_5$[25,26]and $CsCr_3Sb_5$[15], and the Kagome metal $ScV_6Sn_6$[27]. The total DOS at the Fermi level is estimated to be D($E_F$) ~30 $eV^{-1}$ $f.u.^{-1}$, significantly exceeding values typical of conventional Kagome systems. From this, the band-derived Sommerfeld coefficient is calculated as $\gamma_{band} = (\pi^2 k_B^2/3)\, D(E_F) \approx 70$ mJ $mol^{-1}$ $K^{-2}$ per formula unit. The experimental Sommerfeld coefficient greatly exceeds the band value $\gamma_{band}$, indicating pronounced quasiparticle mass renormalization driven by strong electronic correlations.

**2.3 Chemical Modulation**

The electronic ground states of Kagome-related materials often exhibit a remarkable sensitivity to lattice parameters. A prominent example is the $CsCr_3Sb_5$ family, where the application of physical pressure effectively tunes the system from a non-superconducting state into a superconducting one[15]. Inspired by such pressure-induced modulation, we employed chemical pressure to modulate the Mosaic lattice by substituting the $Cs^+$ ions in CVT with smaller $Rb^+$ ions (**Figure 5),** thereby exploring the evolution of the correlated state under structural compression. This ionic substitution leads to a dramatic contraction of the interlayer spacing, with the *c*-axis shrinking from 8.21 Å to 7.93 Å (**Figure 5a**). Concomitantly, the stacking sequence evolves from an A-A-A pattern (space group *P*-62*m*) to a staggered A-B-A-B configuration (*Cmcm*). Although the vanadium layers retain the same Mosaic lattice, on a local scale the vanadium trimers lose their equilateral symmetry and undergo significant distortion, as evidenced by the inequivalent V-V bond lengths in the Rb-analogue (**Figure 5b**). Detailed crystallographic details are provided in **Table S3-4** and **Figure S9-10**. The lattice contraction induced by Rb substitution serves as an effective chemical pressure, which is expected to modify the V 3d orbital overlap and Te-mediated superexchange interactions, thereby providing a route to tune the correlated electronic state in this system.

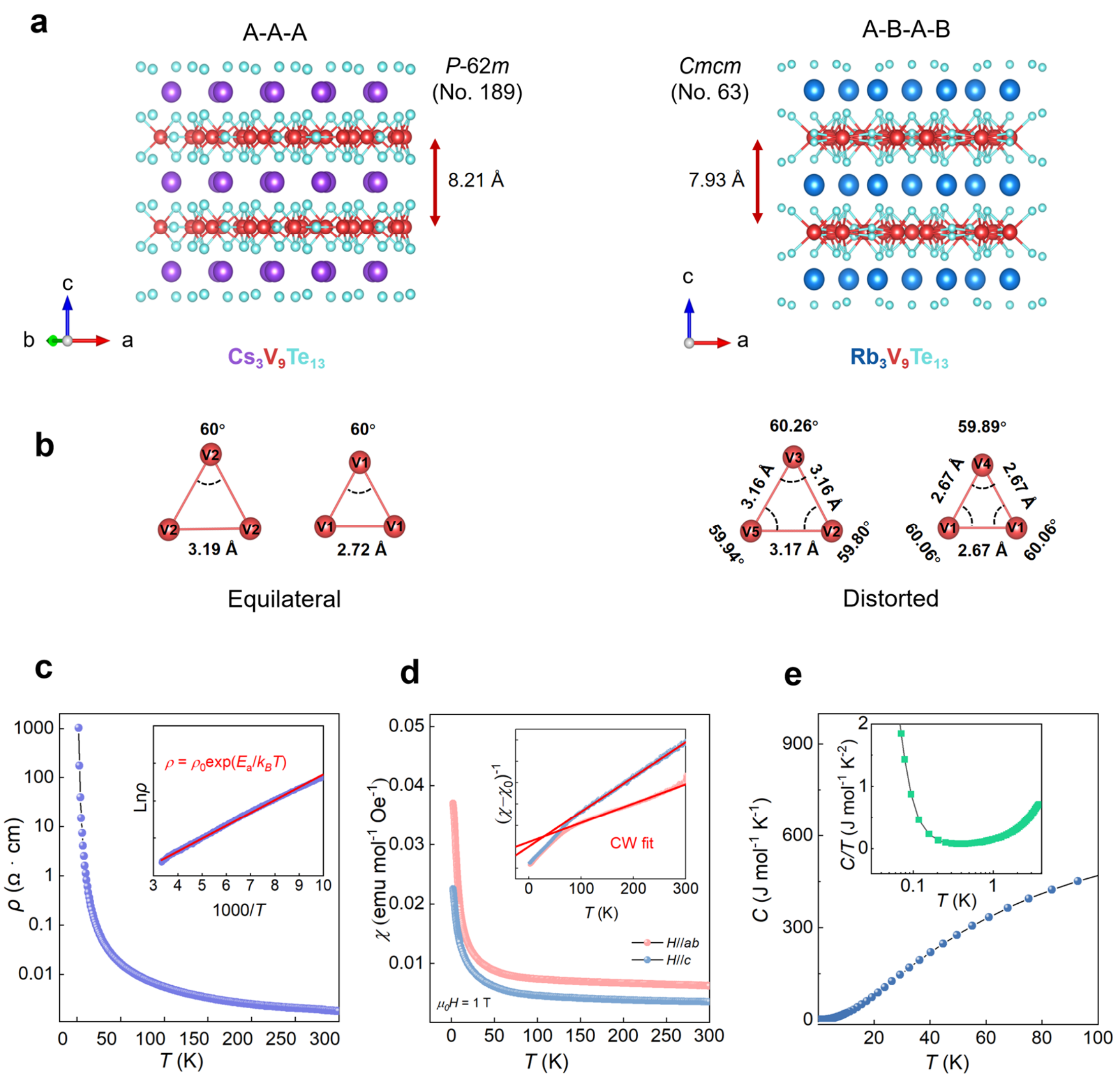

**Figure 5 Structural comparison between CVT and RVT, and physical properties of RVT. a,** The schematic full crystal structure of CVT and RVT. **b**, The equilateral and distorted triangulars in CVT and RVT, respectively. **c**, Temperature dependence of the electrical resistivity for RVT. Inset of the plot of ln($\rho$) against temperature analyzed according to the Arrhenius equation. **d**, Temperature-dependent magnetic susceptibility (ZFC) data for RVT with fields of $\mu_0 H$ = 1 T. Inset of Curie-Weiss fit to the inverse susceptibility data, with the red line representing the fit to $\chi = \chi_0 + C/(T - \theta_W)$. **e**, Temperature-dependent specific heat for RVT. Inset of $C/T$ over the sub-Kelvin temperature range.

## 2.4 Physical Properties of $Rb_3V_9Te_{13}$

The physical properties of $Rb_3V_9Te_{13}$(RVT) diverge sharply from those of CVT. In stark contrast to the metallic CVT, RVT exhibits semiconducting behavior with a substantial activation energy $E_a \approx 15$ meV (**Figure 5c**), indicating a transition toward more localized electronic states. The magnetic susceptibility $\chi(T)$ (**Figure 5d**) reveals no sharp phase transition down to 1.8 K. The Curie-Weiss fit yields negative Weiss temperatures $\theta_{ab}$ = -136 K and $\theta_c$ = -62 K, with effective moments $\mu_{eff}$ = 0.70 $\mu_B$/V. and 0.52 $\mu_B$/V for the *ab*-plane and *c*-axis directions, respectively. These values indicate dominant antiferromagnetic interactions and intrinsic magnetic frustration. At low temperature, the $\chi(T)$ curve exhibits a broad shoulder near ~4 K that is gradually suppressed by increasing magnetic field, indicating the emergence of short-range spin correlations (**Figure S11a**). Consistently, nonlinear $M(H)$ behavior is observed at 1.8 K, whereas at 100 K, linear $M(H)$ behavior without saturation up to 9 T is observed, and no hysteresis is detected at any temperature (**Figure S11b**). The absence of long-range order is further confirmed by heat capacity measurements (**Figure 5e**), which show no $\lambda$-like anomaly down to 60 mK. Suppose the antiferromagnetic transition occurs below the measurement limit; the frustration parameter, $f = |\theta|/T_N$, can be estimated to be on the order of $10^3$, suggesting a highly frustrated system. Such features are characteristic of geometrically frustrated lattices, where competing interactions suppress long-range ordering and stabilize persistent short-range spin correlations[28,29]. Notably, $C/T$ rapidly approaches zero at low temperatures, in stark contrast to the large electronic specific heat coefficient observed in CVT (the upturn below 0.2 K is attributed to a nuclear Schottky anomaly). Collectively, these results demonstrate that chemical pressure induced by Rb substitution contracts the interlayer lattice spacing, significantly suppresses the itinerant correlated quasiparticle state of the Mosaic lattice, and ultimately drives the system into a highly frustrated magnetic semiconducting ground state without long-range magnetic order down to 60 mK.

### 3. Conclusion and Outlook

In summary, we have synthesized and characterized a new family of vanadium-based compounds, $A_3V_9Te_{13}$ ($A$ = Cs, Rb), which features a unique Kagome-derived Mosaic lattice geometry constructed from an ordered tessellation of triangles, squares, and pentagonal structural units. Our investigation reveals that $Cs_3V_9Te_{13}$ hosts a prominent bulk phase transition at $T^* \approx 47$ K, characterized by an itinerant-to-correlated crossover and substantial Fermi surface reconstruction, likely associated with a charge- or spin-related electronic instability. This transition is accompanied by pronounced quasiparticle mass renormalization, as indicated by the exceptionally large Sommerfeld coefficient, uncovering extraordinarily strong electronic correlations in this purely *d*-electron system. The enhanced effective magnetic moment in the Cs system, compared with the nearly nonmagnetic conventional Kagome $AV_3Sb_5$ compounds, further highlights the robust correlated electronic nature inherent to the Mosaic framework.

Furthermore, we demonstrate that the electronic properties of this Kagome-derived Mosaic lattice are highly tunable via chemical pressure. Substituting Cs with Rb systematically suppresses the $T^*$ transition and quenches the strongly correlated quasiparticle response. Upon complete suppression of the electronic ordered state, the system evolves into a semiconducting ground state without long-range magnetic order but with dominant antiferromagnetic interactions. The Rb analogue exhibits no long-range magnetic order down to 60 mK. Collectively, the $A_3V_9Te_{13}$ family provides a versatile platform for exploring the interplay between geometric frustration and strong electron correlations.

The discovery of the $A_3V_9Te_{13}$ family opens several promising avenues for future research. A key priority among these is to further clarify whether the $T^* \approx 47$ K transition arises from electronic or spin correlations in $Cs_3V_9Te_{13}$, for which variable-temperature transmission electron microscopy (TEM), Scanning Tunneling Microscope (STM), and nuclear magnetic resonance (NMR) would be valuable. Given the extreme structural sensitivity of the electronic states demonstrated here, physical pressure or uniaxial strain experiments are highly warranted. Such studies may reveal pressure-induced phase transitions, including the possible emergence of unconventional superconductivity, as observed in related vanadium- or chromium-based Kagome systems[15,30]. On the other hand, the study of heavy-fermion-like behavior in $d$-electron materials is still at an early stage. Quantum oscillation and angle-resolved photoemission spectroscopy (ARPES) measurements would be invaluable to directly probe the effective mass enhancement and detailed band structure, respectively. Furthermore, for $Rb_3V_9Te_{13}$ analogue, microscopic probes such as muon spin rotation/relaxation ($\mu$SR) and inelastic neutron scattering (INS) will be essential to characterize the low-energy spin dynamics and resolve the nature of its frustrated magnetic ground state. Finally, comprehensive theoretical studies focusing on the flat-band physics and orbital-selective correlations of the Mosaic lattice are needed. Unraveling how the unique trimer-based geometry modulates the competition between localized and itinerant electronic degrees of freedom will be key to understanding the rich phase diagram of this new family of materials.

## 4. Method

### 4.1 Single Crystal Growth

Single crystals of $Cs_3V_9Te_{13}$ or $Rb_3V_9Te_{13}$ were grown using a flux method. Starting materials, including Cs or Rb pieces (2N), V powder (3N), and Te lumps (5N) with a total mass of 3 g, were weighed according to a molar ratio of (Cs or Rb) : V : Te = 7 : 3 : 12. The mixture was then loaded into an alumina crucible, which was subsequently sealed in a tantalum tube under an argon atmosphere. The tantalum tube was finally encapsulated in an evacuated quartz ampoule. The assembly was heated to 1000 °C over 15 hours, held at this temperature for 24 hours, and then slowly cooled to 650 °C at a rate of 2 K/h. After the reaction, the excess Te flux was removed by high-temperature centrifugation, followed by mechanical exfoliation and rinsing with deionized water, yielding high-quality single crystals with clean surfaces for subsequent measurements. The polycrystalline sample was synthesized using a procedure similar to that for the single crystal. Stoichiometric amounts of the constituent elements were mixed according to the chemical formula, slowly heated to 900 °C, and annealed for 48 h to obtain the polycrystalline powder. All sample handling procedures were performed inside an argon-filled glovebox ($H_2O$ and $O_2$ levels < 0.1 ppm).

**4.2 Single Crystal X-ray Diffraction (SC-XRD)**

The single crystal X-ray diffraction (SC-XRD) was performed using a single crystal X-ray diffractometer (D8 VENTURE, Bruker) under both 300 K and 100 K with a $N_2$-based cryo cooling system. The PHOTON III C14 detector captured graphite-monochromatized Mo $K_\alpha$ radiation with a wavelength of 0.71073 Å. At 30 K, the SC-XRD pattern was collected using a PHOTON II diffractometer with a multilayer mirror monochromatized Mo Kα radiation ($\lambda$ = 0.71073 Å) and a He-based cryocooling system. APEX4 software was utilized to correct the raw data. The ShelXT structure solution program was used to solve the structures using direct methods, and the ShelXL least-squares refinement package in the Olex2 program was employed to refine the structure[31]. The PLATON program's ADDSYM algorithm was applied to detect any possible higher symmetry, but no missing symmetry was found for the above materials[32]. The related crystallographic data are summarized in **Tables S1-S4**.

**4.3 Powder X-ray diffraction (P-XRD)**

Powder XRD data were collected at room temperature on a Bruker D8 Advance Eco diffractometer using Cu-$K_\alpha$ radiation ($\lambda$ = 1.5418 Å) over a 2θ range of 10° to 90° (**Figures S1 and S9**).

**4.4 X-ray photoelectron spectroscopy (XPS)**

XPS measurements were performed using a Thermo K-Alpha spectrometer. Samples were prepared, cleaved, and handled entirely in an argon-filled glovebox to avoid air exposure, and then transferred to the instrument under inert conditions.

**4.5 Physical Property Measurement System (PPMS)**

Magnetic, transport, and thermal properties were characterized using a Quantum Design PPMS. DC magnetic susceptibilities were measured on single crystals of $Cs_3V_9Te_{13}$ and $Rb_3V_9Te_{13}$ from 1.8 to 300 K under various magnetic fields. Electrical resistivity within the *ab* plane was determined by the standard four-probe method. The Hall coefficient was determined by sweeping the magnetic field from positive to negative at a fixed temperature, with the magnetoresistance component subtracted. Specific heat capacity was measured from 300 down to 1.8 K using the relaxation technique. Measurements were extended to 0.06 K using a dilution refrigerator attachment.

**4.6 Theoretic simulations**

The electronic structure of $Cs_3V_9Te_{13}$ was calculated using density functional theory (DFT) as implemented in the Vienna ab initio Simulation Package (VASP)[33]. The experimentally determined crystal structure was fully relaxed until the residual forces on each atom were smaller than 5 meV/Å. Van der Waals interactions were included using the zero-damping Grimme correction[34] during structural optimization. The ground-state magnetic order is identified as A-type antiferromagnetism, with the V spin moments aligned ferromagnetically within the *ab* plane and antiferromagnetically along the out-of-plane direction. This magnetic configuration is therefore used in the electronic structure calculations. The exchange–correlation potential was treated within the generalized gradient approximation PBE[35] in the structure relaxation and meta-generalized-gradient approximation SCAN[36] in band structure calculation. A plane-wave energy cutoff of 300 eV was employed, and the Brillouin zone was sampled using a 5×5×7 k-point mesh. The spin-orbit coupling is omitted throughout.

**Author contributions**

R.Z. and S.G. conceived and supervised the project, and acquired funding. Y.X., Q.D., X.X., synthesized the samples and performed comprehensive physical property measurements. Z.Q., Z.L., and S.G. conducted structural analysis of the samples. H.T. performed the DFT calculations. All authors participated in data analysis and manuscript preparation.

**Supporting Information**

Analytical data for all new compounds, including crystallographic data, powder X-ray diffraction, X-ray Photoelectron Spectroscopy, and physical property measurements.

The CCDC numbers 2501463 and 2501462 contain the supplementary crystallographic data for $Rb_3V_9Te_{13}$ and $Cs_3V_9Te_{13}$, respectively. These data can be obtained free of charge from The Cambridge Crystallographic Data Centre via www.ccdc.cam.ac.uk/data_request/cif.

**Acknowledgments**

We thank Chi-Ming Yim, Rong Yu, and Baiqing Lv for insightful discussions and valuable technical support in characterization. This work was supported by the National Key R&D of China under 2022YFA1402702 and 2021YFA1401600, National Natural Science Foundation of China with Grants Nos. 12334008, and 12374148. S. G. acknowledges the financial support from the National Natural Science Foundation of China (22205091), the Guangdong Pearl River Talent Plan (2023QN10C793). H.T. is supported by the National Natural Science Foundation of China (12574270) and Science and Technology Commission of Shanghai Municipality (No. 24PJA051). Y.X acknowledges the support of the China Postdoctoral Science Foundation (2026M793724).

**Conflict of Interest**

The authors declare no conflict of interest.

**Data availability**

The data that support the findings of this study are available from the corresponding author upon reasonable request.

# Supporting Information

**Table S1.** Crystal Data and Structure Refinements for $Cs_3V_9Te_{13}$ (CVT) at 30, 100 and 300 K.

| Empirical formula | $Cs_3V_9Te_{13}$ | | |
|---|---|---|---|
| Formula weight | 2516.09 | 2516.09 | 2516.09 |
| Temperature/K | 30 | 100 | 300 |
| Crystal system | Hexagonal | Hexagonal | Hexagonal |
| Space group | *P*-62*m* (189) | *P*-62*m* (189) | *P*-62*m* (189) |
| *a*/Å | 10.1371(5) | 10.1217(5) | 10.089(3) |
| *c*/Å | 8.1465(5) | 8.1481(7) | 8.212(3) |
| $\gamma$ [°] | 120 | 120 | 120 |
| volume/Å$^3$ | 724.98(8) | 722.93(9) | 723.9(5) |
| Z | 1 | 1 | 1 |
| $\rho_{calc}$ /g/cm$^3$ | 5.763 | 5.779 | 5.771 |
| μ/mm$^{-1}$ | 19.286 | 19.341 | 19.315 |
| F(000) | 1048 | 1048 | 1048 |
| Mo Kα radiation/Å | 0.71073 | 0.71073 | 0.71073 |
| 2θ range for data collection/° | 4.64 to 52.99 | 4.65 to 53.94 | 4.66 to 54.95 |
| Goodness-of-fit | 1.188 | 1.142 | 1.104 |
| Largest diff peak and hole (e/A$^3$) | 2.42 and -1.52 | 1.88 and -0.61 | 1.77 and −1.17 |
| Final R indexes [I ≥ 2σ(I)] | $R_1$ = 0.0309, w$R_2$ = 0.0869 | $R_1$ = 0.0180, w$R_2$ = 0.0420 | $R_1$ = 0.0294, w$R_2$ = 0.0685 |
| Final R indexes [all data] | $R_1$ = 0.0314, w$R_2$ = 0.0872 | $R_1$ = 0.0189, w$R_2$ = 0.0427 | $R_1$ = 0.0304, w$R_2$ = 0.0692 |

**Table S2.** Fractional atomic coordinates and equivalent isotropic displacement parameters (Å$^2$) or CVT at 300 K.

| Atom | x | y | z | $U_{eq}$ |
|---|---|---|---|---|
| Cs1 | 0.36401(19) | 0.36401(19) | 0 | 0.0336(4) |
| Te1 | 2/3 | 1/3 | 0.25265(16) | 0.0194(3) |
| Te2 | 0 | 0.22819(10) | 0.71992(13) | 0.0168(3) |
| Te3 | 0.41185(15) | 0.41185(15) | 1/2 | 0.0343(5) |
| V1 | 0.1555(4) | 0.1555(4) | 1/2 | 0.0151(8) |
| V2 | 0.8210(3) | 0.2866(3) | 1/2 | 0.0172(6) |

**Table S3.** Crystal Data and Structure Refinements for $Rb_3V_9Te_{13}$ (RVT) at 300 K.

| Empirical formula | $Rb_3V_9Te_{13}$ |
|---|---|
| Formula weight | 2373.67 |
| Temperature/K | 298(2) |
| Crystal system | Orthorhombic |
| Space group | *Cmcm* (63) |
| $a$/Å | 17.5489(13) |
| $b$/Å | 10.1022(8) |
| $c$/Å | 15.8691(13) |
| volume/Å$^3$ | 2813.3(4) |
| $Z$ | 4 |
| $\rho_{calc}$/g/cm$^3$ | 5.604 |
| $\mu$/mm$^{-1}$ | 21.211 |
| $F$(000) | 3976.0 |
| Mo Kα radiation/Å | 0.71073 |
| 2θ range for data collection/° | 4.64 to 55.00 |
| Goodness-of-fit on $F^2$ | 1.025 |
| Largest diff peak and hole (e/A$^3$) | 2.148 and -1.519 |
| Final $R$ indexes [$I \geq 2\sigma(I)$] | $R_1$=0.0327, $wR_2$=0.0650 |
| Final $R$ indexes [all data] | $R_1$=0.0439, $wR_2$=0.0701 |

**Table S4.** Fractional atomic coordinates and equivalent isotropic displacement parameters (Å$^2$) for RVT at 300 K.

| Atom | x | y | z | $U_{eq}$ | Occupancy |
|---|---|---|---|---|---|
| Te1 | 0.66640(3) | 0.33796(6) | 0.62023(3) | 0.02561(15) | 1 |
| Te2 | 0.38528(3) | 0.72407(5) | 0.63700(3) | 0.01983(14) | 1 |
| Te3 | 0 | 0.56919(7) | 0.86402(5) | 0.01993(18) | 1 |
| Te4 | 0.79518(6) | 0.54361(10) | 0.77289(14) | 0.0463(9) | 0.5 |
| Te5 | 1/2 | 0.42951(14) | 0.77332(16) | 0.0418(9) | 0.5 |
| Rb1 | 1/2 | 1/2 | 1/2 | 0.0516(6) | 1 |
| Rb2 | 0.82328(9) | 1/2 | 1/2 | 0.0465(4) | 1 |
| V1 | 0.92399(11) | 0.41591(17) | 3/4 | 0.0186(4) | 1 |
| V2 | 0.76591(11) | 0.28358(18) | 3/4 | 0.0188(4) | 1 |
| V3 | 0.64462(11) | 0.51489(18) | 3/4 | 0.0184(4) | 1 |
| V4 | 1/2 | 0.6867(2) | 3/4 | 0.0177(6) | 1 |
| V5 | 0.41074(11) | 0.21761(18) | 3/4 | 0.0184(4) | 1 |

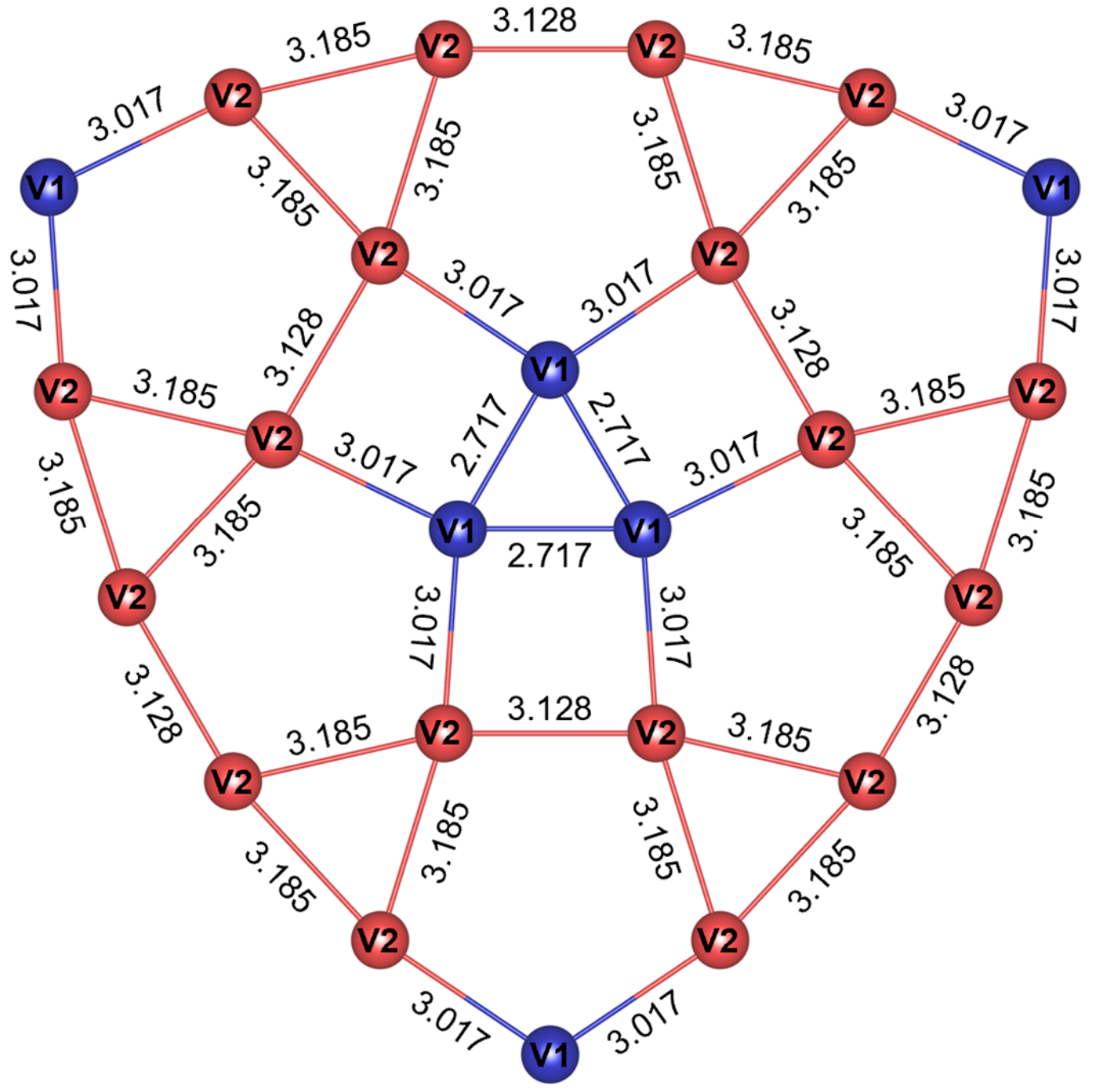

**Figure S1.** Mosaic lattice details for CVT.

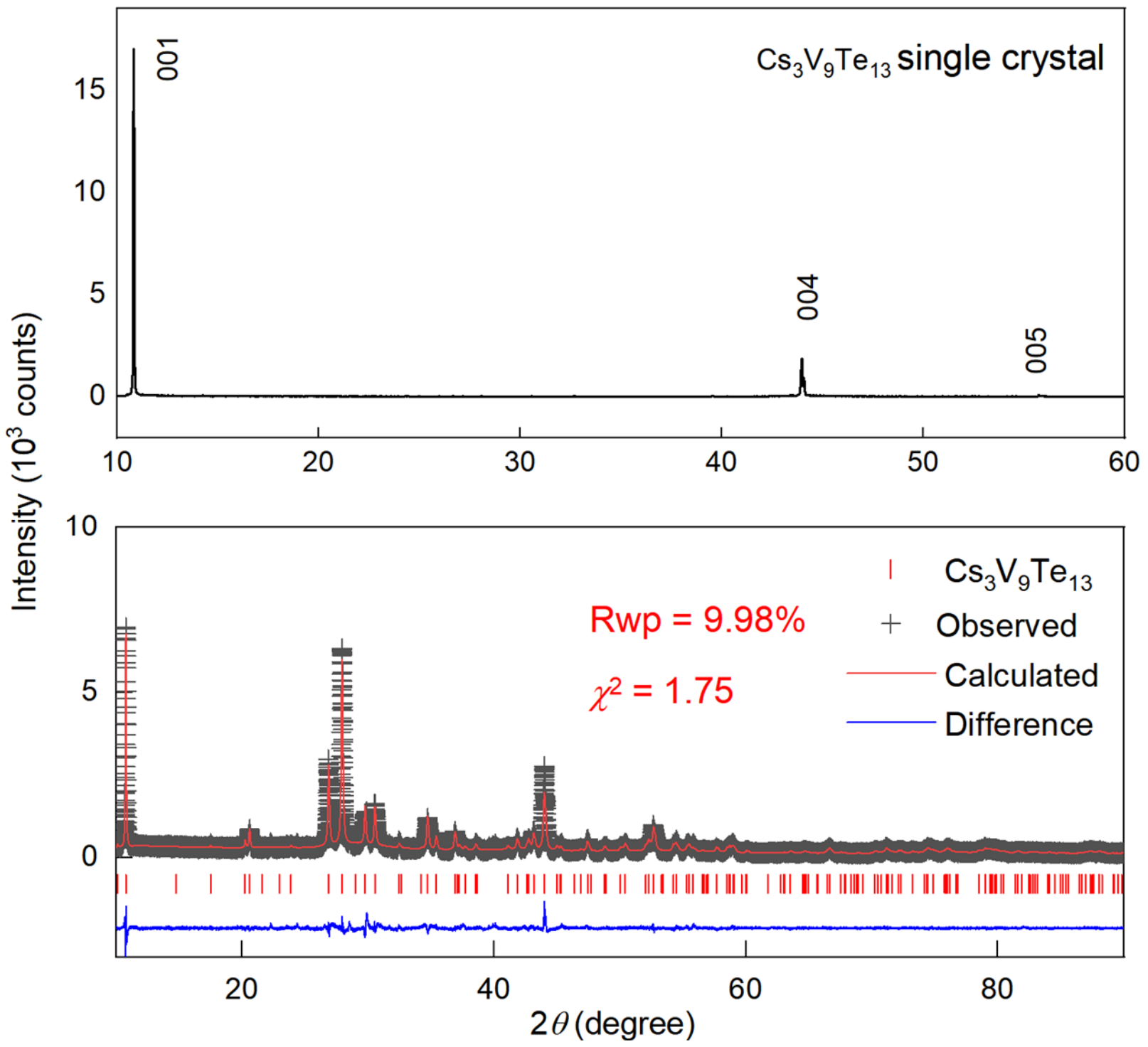

**Figure S2.** X-ray diffraction pattern of the CVT single crystal and Rietveld refinement profile of polycrystalline sample.

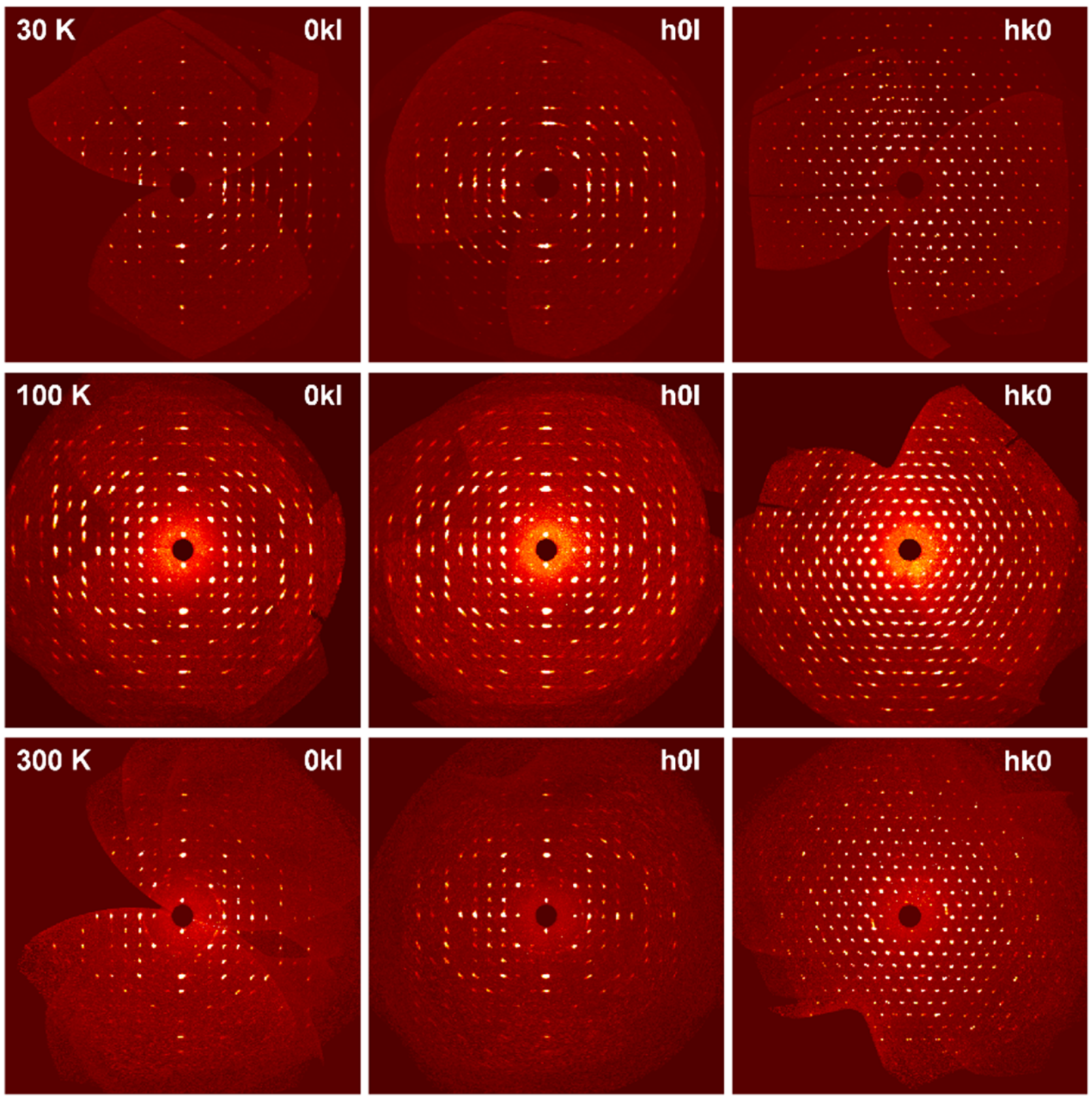

**Figure S3.** Diffracted reflections in reciprocal lattice planes (0kl), (h0l), and (hk0) determined from experimental data at temperatures of 300 K, 100 K, and 30 K.

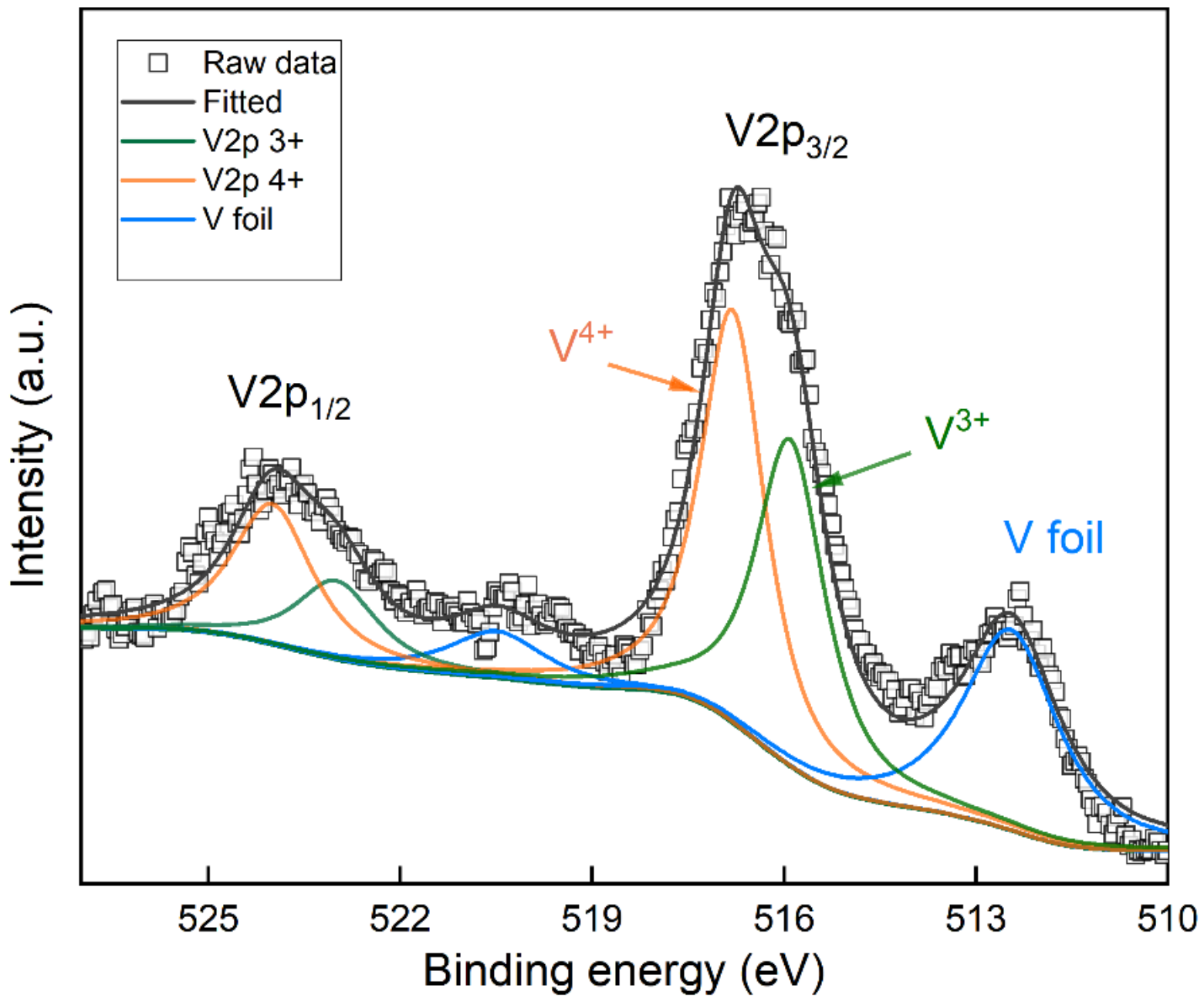

**Figure S4.** X-ray Photoelectron Spectroscopy (XPS) spectra for CVT single crystal.

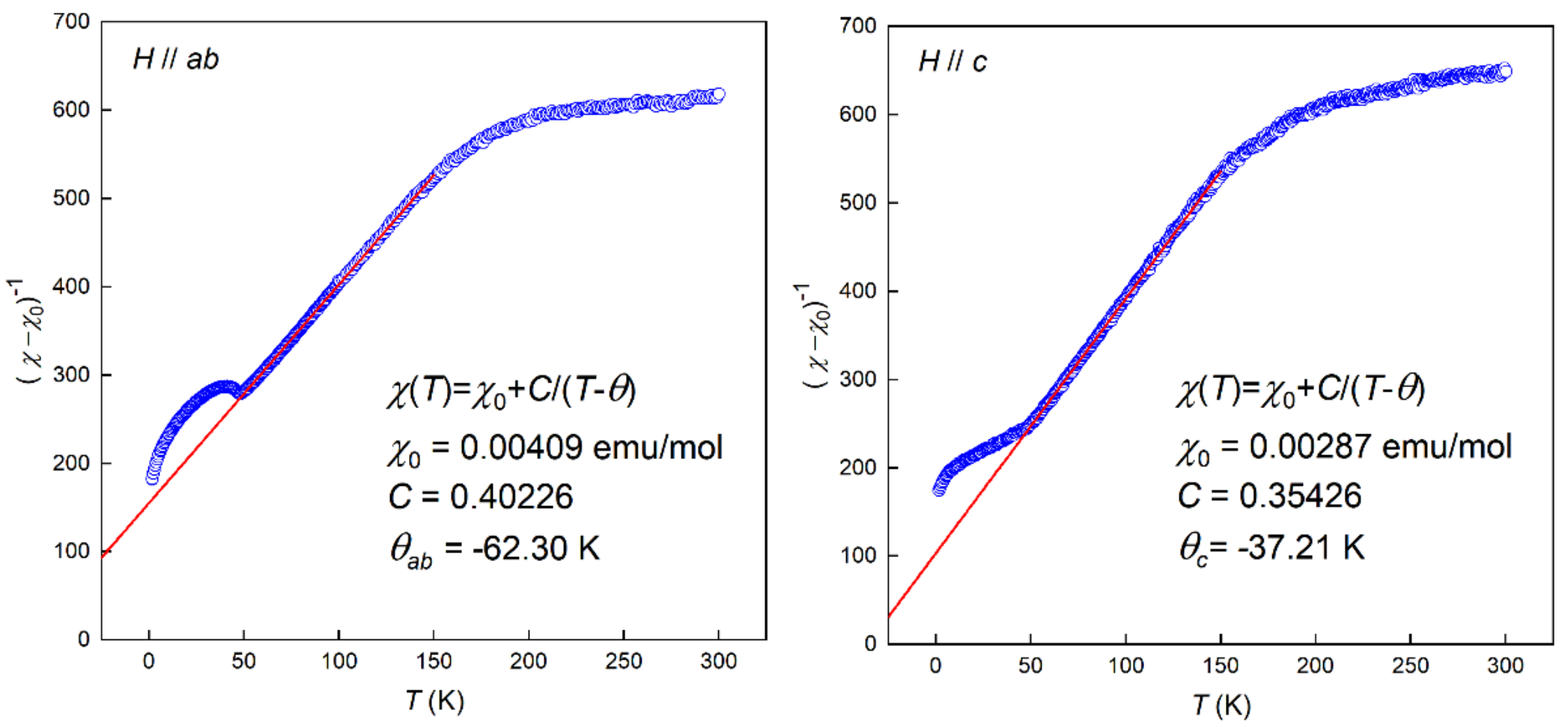

**Figure S5.** Temperature-dependent inverse magnetic susceptibility, $(\chi\text{-}\chi_0)^{-1}$, of CVT measured under $\mu_0 H$ = 1 T with the field applied parallel to the *ab*-plane and the *c*-axis.

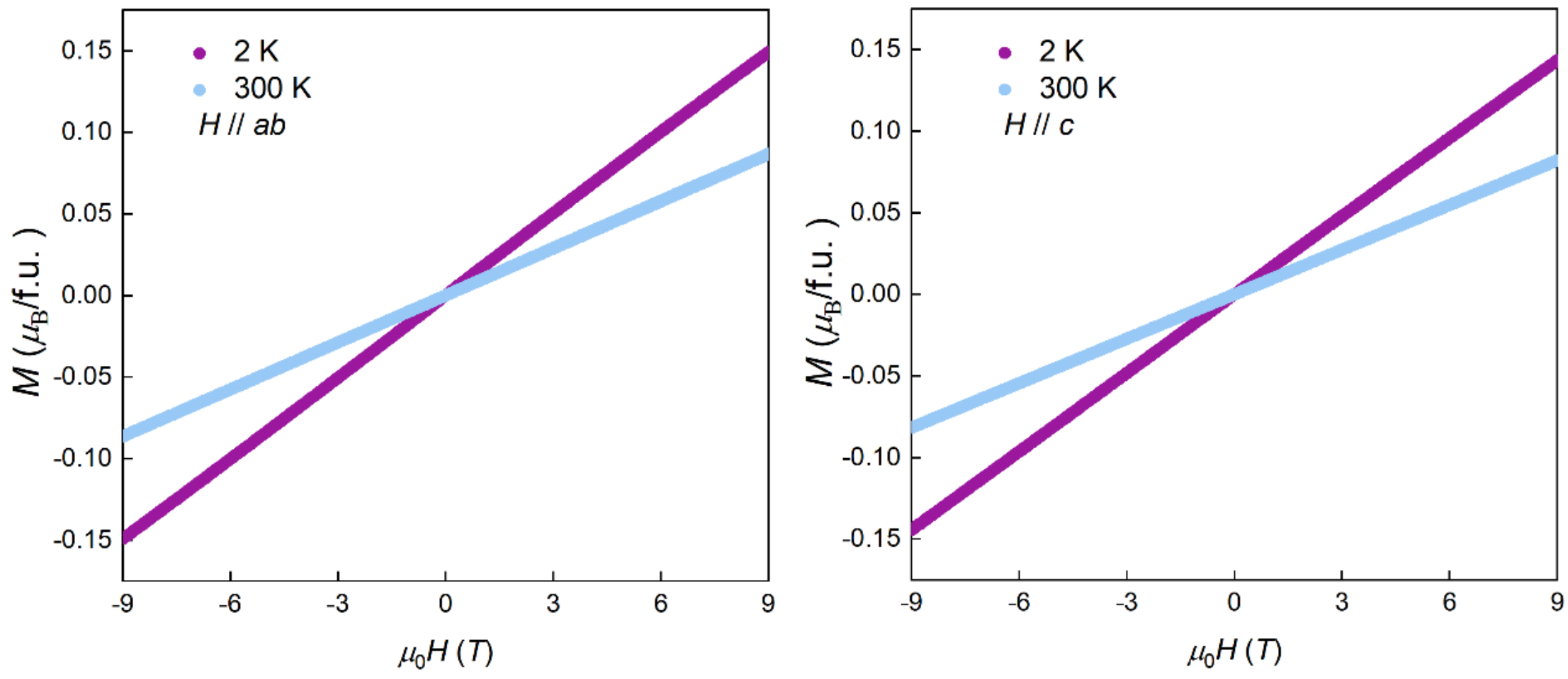

**Figure S6.** Field-dependent magnetization of CVT, measured at various temperatures for the field applied parallel to the *ab*-plane and the *c*-axis.

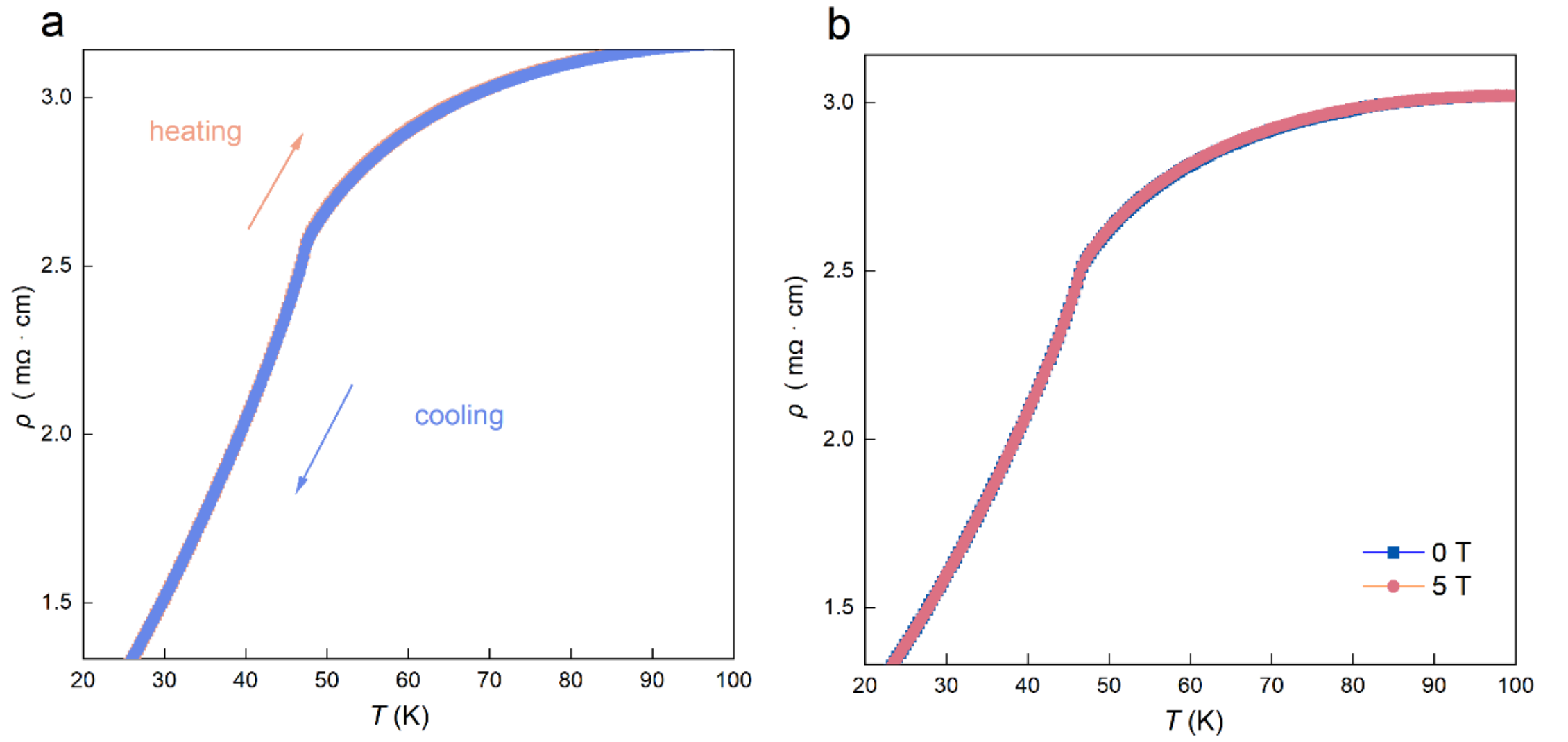

**Figure S7.** **a**, Temperature-dependent resistivity under zero magnetic field during heating and cooling cycles. **b**, Temperature-dependent resistivity under different magnetic fields.

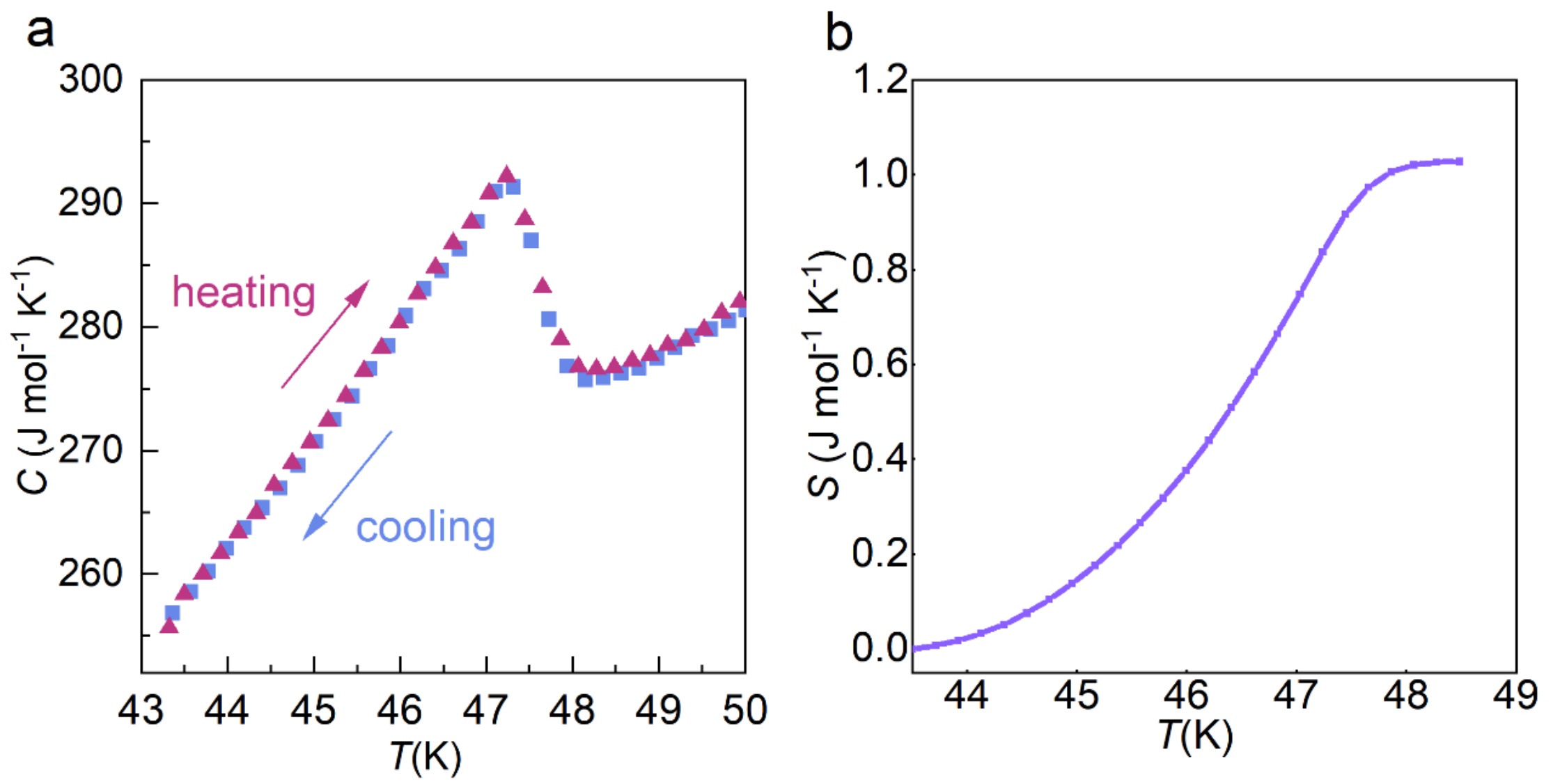

**Figure S8.** **a**, Temperature-dependent specific heat measured during heating and cooling cycles under zero magnetic field. **b**, Temperature dependence of the entropy change extracted by subtracting a polynomial fit background.

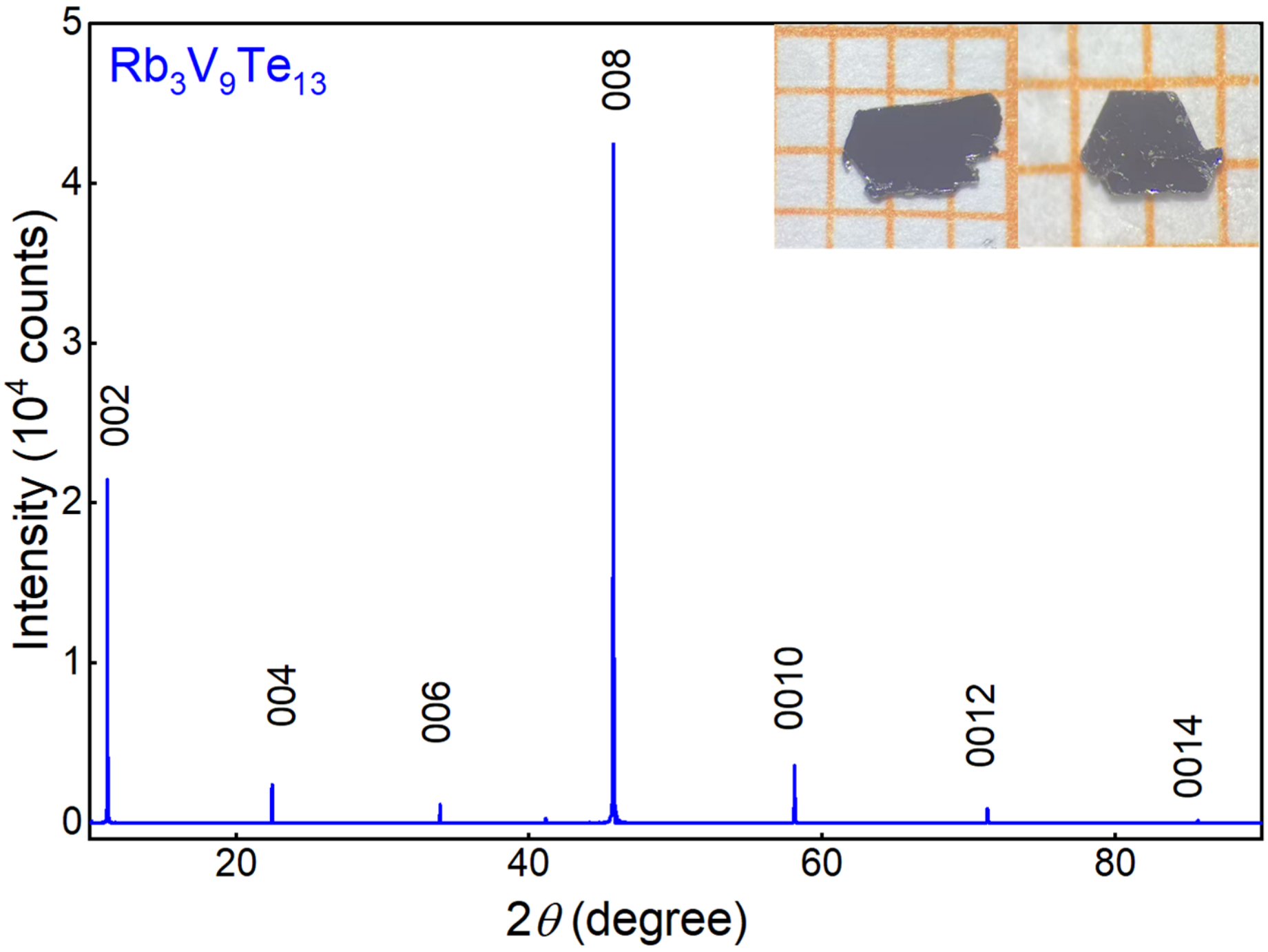

**Figure S9** X-ray diffraction pattern of the RVT single crystal. The inset shows an optical image of the RVT single crystals.

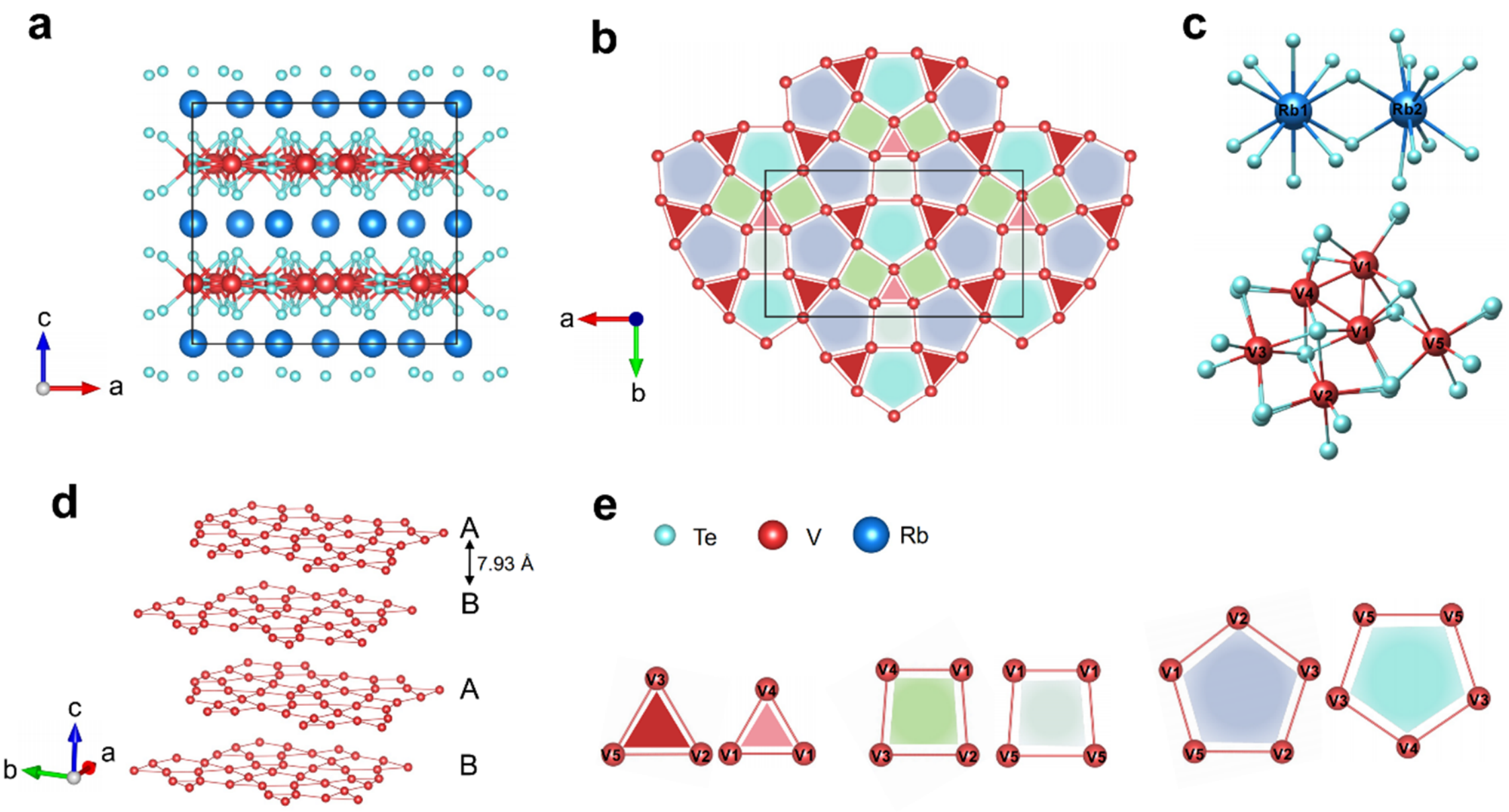

**Figure S10.** Crystal structure of RVT. **a**, 3D framework of RVT. **b**, The V-based 2D Mosaic Lattice. **c**, Coordination environments for Rb and V atoms, respectively. **d**, The stacking form of V-based layers. **e**, Constituent triangular, square, and pentagonal units in a 2D Mosaic Lattice.

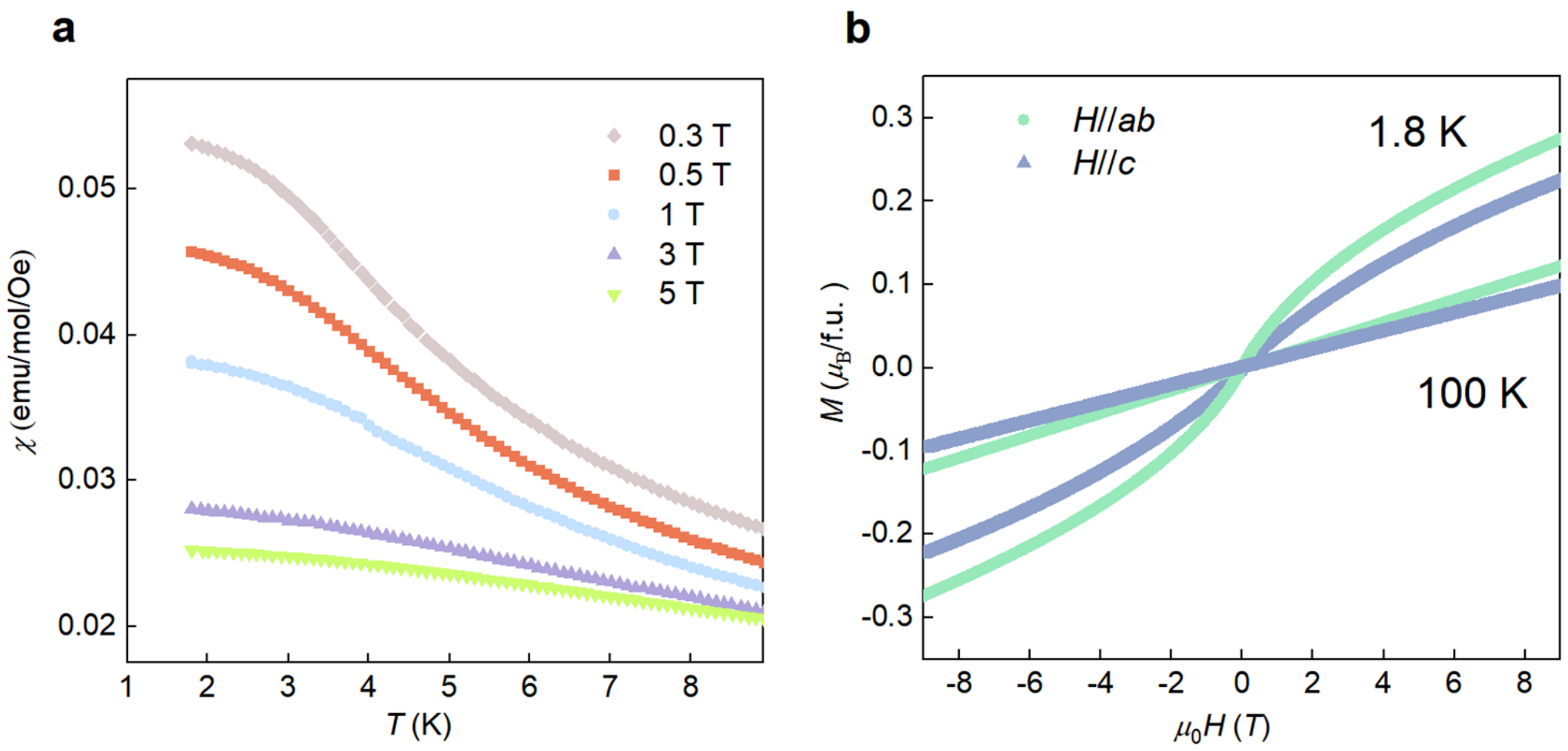

**Figure S11. a,** Temperature-dependent magnetization of RVT, measured under different magnetic fields applied parallel to the *ab*-plane. **b,** Field-dependent magnetization of RVT at various temperatures for the field applied parallel to the *ab*-plane and the *c*-axis.